\documentclass[letterpaper, 10 pt, journal, twoside]{IEEEtran}
\usepackage{graphicx}
\usepackage{epstopdf}
\usepackage{amsmath, amssymb, psfrag}
\usepackage{graphicx,paralist}
\usepackage{mathtools}
\usepackage{threeparttable, balance}

\usepackage{epsfig} 
\usepackage{mathptmx} 
\usepackage{times} 
\usepackage{float}
\usepackage{cite}
\usepackage{algorithm}
\usepackage{hyperref, url}
\usepackage{algpseudocode}
\usepackage{algorithm}
\usepackage{soul, physics}
\usepackage{amsthm}

\newtheorem{remark}{Remark}

\ifCLASSINFOpdf
\else
\fi
\begin{document}

\title{Non-Prehensile Manipulation of a Devil-Stick:\\
Planar Symmetric Juggling Using Impulsive Forces}

\author{Nilay Kant$^{1}$, Ranjan Mukherjee$^{1}$%
\thanks{
This work was supported by the National Science Foundation, NSF Grant CMMI-1462118.}
\thanks{$^{1}$Nilay Kant and Ranjan Mukherjee are with Mechanical Engg. Dept., Michigan State University, MI 48824, USA,
{\tt\footnotesize mukherji@egr.msu.edu}}
\vspace{-0.05in}
}

\maketitle

\begin{abstract}

Juggling a devil-stick can be described as a problem of non-prehensile manipulation. Assuming that the devil-stick remains confined to the vertical plane, the problem of juggling the stick between two symmetric configurations is considered. Impulsive forces are applied to the stick intermittently and the impulse of the force and its point of application are modeled as control inputs to the system. The dynamics of the devil-stick due to the impulsive forces and gravity is described by half-return maps between two Poincar\'e sections; the symmetric configurations are fixed points of these sections. A coordinate transformation is used to convert the juggling problem to that of stabilization of one of the fixed points. Inclusion of the coordinate transformation in the dynamic model results in a nonlinear discrete-time system. A dead-beat design for one of the inputs simplifies the control problem and results in a linear time-invariant discrete-time system. Standard control techniques are used to show that symmetric juggling can be achieved from arbitrary initial conditions.

\end{abstract}


\IEEEpeerreviewmaketitle

\section{Introduction} \label{sec1}
A devil-stick is typically juggled using two hand sticks and several tricks can be performed depending on the proficiency of the juggler. Some of the common tricks are: \emph{standard-idle}, \emph{flip-idle}, \emph{airplane-spin} or \emph{propeller}, \emph{top-only idle}, and \emph{helicopter} \cite{youtube1}. The \emph{top-only idle} is one of the simplest tricks and is the focus of this investigation; a video tutorial for learning this trick can be found in \cite{youtube2}. In top-only idle, intermittent forces are applied to the devil-stick. Since the devil-stick is never grasped, the juggling problem can be viewed as a non-prehensile manipulation problem. If robots are to perform this trick, the motion of the end-effectors would have to be coordinated and controlled to apply the correct magnitude of impulsive forces to the devil-stick at appropriate locations. We do not focus on the end-effector motion control problem (see \cite{hirai2006dynamic, kober2012playing} for application to ball juggling); instead, we investigate the magnitude and location of the forces needed to perform the \emph{top-only idle} trick.\

Many juggling tasks, including the \emph{top-only idle} trick, involve intermittent application of impulsive forces and several researchers \cite{tornambe1999modeling, lynch2001recurrence, zavala2001direct, brogliato2006controllability, goedel2012hybrid} have studied the controllability and stability of such systems. Although impulsive control of the devil-stick has not been investigated, the control problem associated with juggling of balls and air-hockey pucks has seen several solutions \cite{ronsse2007rhythmic, spong2001impact, schaal1993open, sanfelice2007hybrid}. In all of these solutions, the object being juggled has been modeled as a point mass and its orientation is excluded from the dynamic model. In contrast, for devil-stick tricks such as \emph{top-only idle}, the stick is shuffled between two symmetric configurations about the vertical; therefore, the orientation of the stick must be included in the dynamic model.\

In earlier works on the devil-stick \cite{nakaura2004enduring, shiriaev2006generating}, controllers have been designed for \emph{airplane-spin} or \emph{propeller}-type motion; a single hand-stick is used to rotate the devil-stick about a virtual horizontal axis using continuous-time inputs. The dynamics model and control design of \emph{top-only idle} motion of the devil-stick has not appeared in the literature; to the best of our knowledge, it is presented here for the first time. It is assumed that impulsive forces are applied intermittently to the devil-stick and the control inputs are the impulse of the force and its point of application on the stick. The control inputs are designed to juggle the stick between two symmetric configurations about the vertical, starting from an arbitrary initial configuration.\

This paper is organized as follows. The juggling problem is formally described in section \ref{sec2}. The dynamics of the devil-stick is presented in section \ref{sec3}; it is comprised of impulsive dynamics due to the control inputs and continuous dynamics due to torque-free motion under gravity. A coordinate transformation is used to simplify the control problem and the dynamics is described by a nonlinear discrete-time system. The control design is provided in section \ref{sec4}. By choosing one of the control inputs to be dead-beat, the nonlinear system is simplified to a linear discrete-time system. For stable juggling, the linear system is controlled using linear quadratic regulator (LQR) and model predictive control (MPC) techniques. Simulation results are presented in section \ref{sec5} and concluding remarks are provided in section \ref{sec6}.\

\section{Problem Description}\label{sec2}
\begin{figure}[b!]
\centering
\psfrag{A}[][]{{$x$}}
\psfrag{B}[][]{{$y$}}
\psfrag{C}[][]{{$\ell$}}
\psfrag{D}[][]{{$\theta$}}
\psfrag{E}[][]{{$m, J$}}
\psfrag{F}[][]{\small{$G \equiv (h_x, h_y)$}}
\psfrag{K}[][]{\small{$g$}}
\includegraphics[width=0.46\hsize]{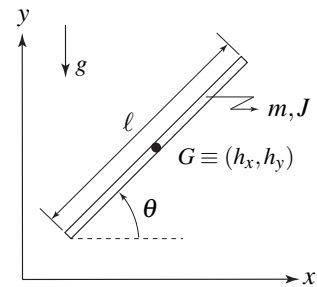}
\caption{A three degree-of-freedom of a devil-stick.}
\label{Fig1}
\end{figure}
\begin{figure}[t!]
\centering
\psfrag{A}[][]{{$x$}}
\psfrag{B}[][]{{$y$}}
\psfrag{C}[][]{\small{$\pi-\theta^*$}}
\psfrag{D}[][]{\small{$\theta^*$}}
\psfrag{E}[][]{\small{$(-h_x^*, h_y^*)$}}
\psfrag{F}[][]{\small{$(h_x^*, h_y^*)$}}
\psfrag{K}[][]{\small{$g$}}
\psfrag{P}[][]{\small{$I = I^*$}}
\psfrag{Q}[][]{\small{$r = r^*$}}
\includegraphics[width=0.91\hsize]{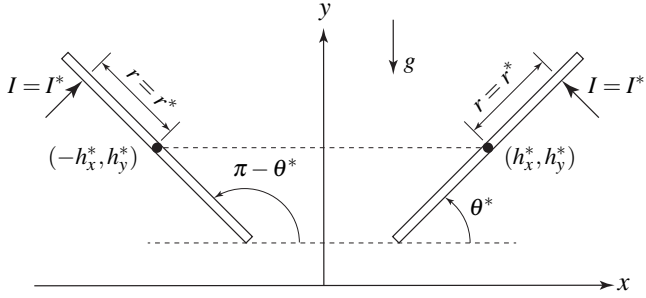}
\caption{Symmetric configurations of the devil-stick in Fig.\ref{Fig1}.}
\label{Fig2}
\end{figure}
Consider the three degree-of-freedom devil-stick shown in Fig. \ref{Fig1}, which can move freely in the  $xy$ vertical plane. The stick has length $\ell$, mass $m$, and mass moment of inertia $J$ about its center-of-mass G. The configuration of the stick is described by the three generalized coordinates: $(\theta, h_x, h_y)$, where $\theta$ is the orientation of the stick with respect to the positive $x$ axis, measured counter-clockwise, and $(h_x, h_y)$ are the Cartesian coordinates of G. The objective is to juggle the stick between two configurations that are symmetric with respect to the vertical axis. The coordinates of the stick in these two configurations are $(\theta^*, h_x^*, h_y^*)$ and $(\pi - \theta^*, -h_x^*, h_y^*)$, where $\theta^* \in (0, \pi/2)$ - see Fig.\ref{Fig2}. It is assumed that juggling is achieved by applying impulsive forces perpendicular to the stick; they are applied only when the orientation of the stick is $\theta = \theta^*$ or $\theta = \pi - \theta^*$. Therefore, the time of application of the impulsive force is not a part of the control design. The control inputs are the pair $(I, r)$, where $I$, $I \geq 0$, is the impulse of the impulsive force and $r$ is the distance of the point of application of the force from G. The value of $r$ is considered to be positive if the angular impulse of the impulsive force about G is in the positive $z$ direction when $\theta=\theta^*$, and is in the negative $z$ direction when $\theta=\pi-\theta^*$. The control inputs that juggle the stick between the symmetric configurations are denoted by the pair $(I^*, r^*)$.\

\section{Dynamics of the Devil-Stick}\label{sec3}
\subsection{Impulsive Dynamics}\label{sec3-1}
The dynamics of the three-DOF devil-stick is described by the six-dimensional state vector $X$, where
\begin{align*}
X = \left[\begin{matrix} \theta &\omega &h_x &v_x &h_y &v_y \end{matrix}\right]^T, \,\, \omega \triangleq \dot\theta, \,\, v_x \triangleq \dot h_x, \,\, v_y \triangleq \dot h_y
\end{align*}
\noindent Let $t_k$, $k = 1, 2, 3, \cdots$, denote the instants of time when the impulsive inputs are applied. Furthermore, without loss of generality, let $k = (2n-1)$, $n = 1, 2, \cdots$ denote the instants of time when the impulsive inputs are applied at $\theta = \theta^*$, and $k = 2n$, $n = 1, 2, \cdots$ denote the instants of time when the impulsive inputs are applied at $\theta = \pi-\theta^*$. If $t_k^{-}$ and $t_k^{+}$ denote the instants of time immediately before and after application of the impulsive inputs, the linear and angular impulse-momentum relationships can be used to describe the impulsive dynamics\footnote{Impulsive inputs cause discontinuous jumps in the velocity coordinates but no change in the position coordinates. The dynamics of underactuated systems subjected to impulsive inputs is discussed in \cite{mathis2014impulsive,kant, kant2018impulsive, 8814463}.} as follows, for $k = 1, 3, 5, \cdots$
\begin{equation}
X(t_k^{+}) = X(t_k^{-}) + \left[\begin{matrix} 0 \cr (I_k\, r_k/J) \cr 0 \cr-(I_k/m)\sin\theta^* \cr 0 \cr \,\,\,(I_k/m)\cos\theta^*\end{matrix}\right] \label{eq1}
\end{equation}
\noindent and for $k = 2, 4, 6, \cdots$
\begin{equation}
X(t_k^{+}) = X(t_k^{-}) + \left[\begin{matrix} 0 \cr -(I_k\, r_k/J) \cr 0 \cr (I_k/m)\sin\theta^* \cr 0 \cr (I_k/m)\cos\theta^*\end{matrix}\right] \label{eq2}
\end{equation}
\noindent where $(I_k, r_k)$ denote the control inputs at time $t_k$. Between two consecutive impulsive inputs, the devil-stick undergoes torque-free motion under gravity; this is discussed next.\

\subsection{Continuous-time Dynamics}\label{sec3-2}
Over the interval $t \in [t_k^{+}, t_{k+1}^{-}]$, the devil-stick will be in flight; its center-of-mass G will undergo projectile motion and its angular momentum will remain conserved. This dynamics is described by the differential equation:
\begin{equation} \label{eq3}
\dot X = \left[\begin{matrix} \omega\, &0\, &v_x\, &0\, &v_y\, &-g\end{matrix}\right]^T
\end{equation}
\noindent where the initial condition $X(t_k^{+})$ can be obtained from (\ref{eq1}) or (\ref{eq2}), depending on whether $k$ is odd or even.\

\subsection{Poincar\'e Sections and Half-Return Maps}\label{sec3-3}
For the hybrid system, described by impulsive dynamics of section \ref{sec3-1} and continuous dynamics of section \ref{sec3-2}, we define two Poincar\'e sections\footnote{Poincar\'e sections have been previously used for design of gaits for bipedal robots \cite{grizzle2001asymptotically, westervelt2003hybrid}.}$^,$\footnote{It is assumed that the initial conditions of the devil-stick are such that its trajectory intersects one of the two Poincar\'e sections before the first impulsive control input is applied.} \cite{wiggins2003introduction} $S_r$ and $S_l$ as follows:
\begin{equation} \label{eq4}
\begin{aligned}
S_r &: \{X \in \mathbb{R}^{6} \,\,\mid\,\, \theta=\theta^*\} \cr
S_l &: \{X \in \mathbb{R}^{6} \,\,\mid\,\, \theta= \pi - \theta^*\}
\end{aligned}
\end{equation}
These Poincar\'e sections are chosen since the impulsive inputs are applied only when $\theta$ is equal to $\theta^*$ or $(\pi-\theta^*)$. Any point on $S_r$ and $S_l$ can be described by the vector $Y$, $Y \subset X$, where
\begin{align}  \label{eq5}
Y = \left[\begin{matrix} \omega &h_x &v_x &h_y &v_y \end{matrix}\right]^T
\end{align}
The map $\mathbb{P}_r : S_r \rightarrow S_l$  can be determined from (\ref{eq1}) and (\ref{eq3}) as follows:
\begin{equation}\label{eq6}
Y(t_{k+1}^{-}) = A\, Y(t_{k}^{-}) + B_r
\end{equation}
\begin{equation*}
A \triangleq \!\left[\begin{matrix} 1 &0 &0 &0 &0 \cr 0 &1 &\delta_k &0 &0 \cr 0 &0 &1 &0 &0 \cr 0 &0 &0 &1 &\delta_k \cr 0 &0 &0 &0 &1\end{matrix}\right]\!,\,
B_r \triangleq \!\left[\begin{matrix} (I_k\, r_k/J) \cr - (I_k/m)\sin\theta^* \delta_k \cr - (I_k/m)\sin\theta^* \cr (I_k/m)\cos\theta^* \delta_k \!- \!(1/2) g\, \delta_k^2 \cr (I_k/m)\cos\theta^*\!- \! g \delta_k \end{matrix}\right]
\end{equation*}
where $\delta_k \triangleq (t_{k+1}^{-} - t_{k}^{-})$ and $k = (2n-1)$, $n = 1, 2, \cdots$. Similarly, the map $\mathbb{P}_l : S_l \rightarrow S_r$ can be determined from (\ref{eq2}) and (\ref{eq3}) as follows
\begin{equation}\label{eq7}
Y(t_{k+1}^{-}) = A\, Y(t_{k}^{-}) + B_l
\end{equation}
\begin{equation*}
B_l \triangleq \!\left[\begin{matrix} -(I_k\, r_k/J) \cr (I_k/m)\sin\theta^* \delta_k \cr (I_k/m)\sin\theta^* \cr (I_k/m)\cos\theta^* \delta_k \!- \!(1/2) g\, \delta_k^2 \cr (I_k/m)\cos\theta^*\!- \! g \delta_k \end{matrix}\right]
\end{equation*}
\noindent where $k = 2n$, $n = 1, 2, \cdots$. Both $\mathbb{P}_r $ and $\mathbb{P}_l$ in (\ref{eq6}) and (\ref{eq7}), respectively,  can be viewed as half-return maps\footnote{Half-return maps have been used to analyze the behavior of dynamical systems such as the van der Pol oscillator \cite{guckenheimer2003forced, bold2003forced}.} since the composition of these maps are the return maps $\mathbb{P}_r \circ \mathbb{P}_l : S_l \rightarrow S_l$ and $\mathbb{P}_l \circ \mathbb{P}_r : S_r \rightarrow S_r$. In the next section we introduce a coordinate transformation to show that the map $\mathbb{P}_l$, in the transformed coordinates, is identical to
$\mathbb{P}_r$. This simplifies the analysis of the problem.\

\subsection{Coordinate Transformation}\label{sec3-4}
\begin{figure}[b!]
\centering
\psfrag{A}[][]{{$x$}}
\psfrag{B}[][]{{$y$}}
\psfrag{C}[][]{{$z$}}
\psfrag{K}[][]{\small{$g$}}
\psfrag{D}[][][1][-15]{\small{$\theta^*$}}
\psfrag{E}[][][1][-15]{\small{$\pi\! - \!\theta^*$}}
\psfrag{F}[][]{\small{$h_x^*$}}
\psfrag{M}[][]{\small{$P$}}
\psfrag{N}[][]{\small{$Q$}}
\psfrag{P}[][][1][-15]{\footnotesize{$(h_x^*, h_y^*)$}}
\psfrag{Q}[][][1][-15]{\footnotesize{$(-h_x^*, h_y^*)$}}
\includegraphics[width=0.91\hsize]{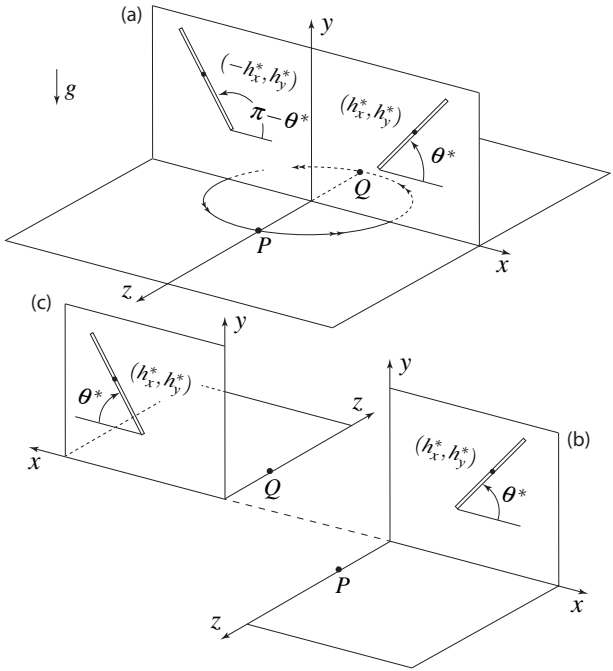}
\caption{(a) Ambidexterous juggler standing at $P$ and applying control actions with both hands, (b) right-handed juggler standing at $P$ and applying control action with right hand, (c) right-handed juggler standing at $Q$ and applying control action with right hand.}
\label{Fig3}
\end{figure}
Consider Fig.\ref{Fig3}, where $z=0$ denotes the $xy$ plane in which the devil-stick is juggled. Typically, the juggler will stand at a point on the positive $z$ axis, denoted by $P$ in Fig.\ref{Fig3} (a), and face the $z=0$ plane. The juggler will apply a control action with the right hand when $\theta=\theta^*$, and with the left hand when $\theta=\pi-\theta^*$, \emph{i.e.}, the juggler is ambidextrous. Instead of alternating between the right and left hands, the juggler can choose to apply all control actions using the right hand only. This juggler, whom we will now refer to as the right-handed juggler, To this end, the juggler will apply the control action standing at $P$ when $\theta=\theta^*$ - see Fig.\ref{Fig3} (b), and apply the next control action after changing location to $Q$ (mirror image of $P$) and facing the $z=0$ plane when $\theta=\pi-\theta^*$ - see Fig.\ref{Fig3} (c). When the devil stick has the orientation $\theta=\pi-\theta^*$, as seen by an observer at $P$, it will have the orientation $\theta=\theta^*$ for the right-handed juggler. After applying control action at $Q$, the right-handed juggler will return back to $P$. If $xyz$ denotes the rotating coordinate frame of the right-handed juggler, the change in position of this juggler can be described by the coordinate transformation:
\begin{align*}
\left[\begin{matrix}x \cr y \cr z \end{matrix}\right]_{\!Q} = R_{y, \pi} \left[\begin{matrix}x \cr y \cr z \end{matrix}\right]_{\!P}, \quad \left[\begin{matrix}x \cr y \cr z \end{matrix}\right]_{\!P} = R_{y, \pi} \left[\begin{matrix}x \cr y \cr z \end{matrix}\right]_{\!Q}
\end{align*}
\noindent where
\begin{align*}
R_{y, \pi} \triangleq {\rm diag}[\begin{matrix} -1, &1, &-1 \end{matrix}]
\end{align*}
\noindent Since $R_{y, \pi}$ changes the sign of the $x$ and $z$ coordinates and leaves the $y$ coordinate unchanged, we can show
\begin{align*}
Y_{\!Q} = R\, Y_{\!P}, \quad Y_{\!P} = R\, Y_{\!Q}
\end{align*}
\noindent where
\begin{align*}
R = R^{-1} \triangleq {\rm diag}[\begin{matrix} -1, &-1, &-1, &1, &1 \end{matrix}]
\end{align*}
\noindent and $Y_{\!P}$ and $Y_{\!Q}$ denote the vector $Y$ as seen by the right-handed juggler standing at points $P$ and $Q$, respectively.\

\subsection{Single Return Map and Discrete-Time Model}\label{sec3-5}
In the reference frame of the right-handed juggler, who alternates between positions $P$ and $Q$, the two Poincar\'e sections $S_l$ and $S_r$ are identical, and equal to
\begin{equation} \label{eq8}
S: \{X \in \mathbb{R}^{6} \,\,\mid\,\, \theta=\theta^*\}
\end{equation}
\noindent This follows from our discussion in section \ref{sec3-4} as well as Figs.\ref{Fig3} (b) and (c). The half-return maps $\mathbb{P}_r$ and $\mathbb{P}_l$ in (\ref{eq6}) and (\ref{eq7}) can be rewritten as follows:
\begin{subequations}\label{eq9}
\begin{align}
Y_{\!P}(t_{k+1}^{-}) &= A\, Y_{\!P}(t_{k}^{-}) + B_r,  \quad k = 1, 3, 5, \cdots \label{eq9a}\\
Y_{\!P}(t_{k+1}^{-}) &= A\, Y_{\!P}(t_{k}^{-}) + B_l,  \quad k = 2, 4, 6, \cdots \label{eq9b}
\end{align}
\end{subequations}
\noindent to explicitly indicate the reference frame of $Y$. Since the right-handed juggler alternates between positions $P$ and $Q$, the half-return map $\mathbb{P}_l$ in (\ref{eq9b}) can be transformed as follows:
\begin{align}\label{eq10}
RY_{\!P}(t_{k+1}^{-}) &= RA\, Y_{\!P}(t_{k}^{-}) + RB_l \cr
\Rightarrow \quad Y_{\!Q}(t_{k+1}^{-}) &= AR\, Y_{\!P}(t_{k}^{-}) + RB_l \cr
\Rightarrow \quad Y_{\!Q}(t_{k+1}^{-}) &= A\, Y_{\!Q}(t_{k}^{-}) + B_r,  \quad k = 2, 4, 6, \cdots\qquad
\end{align}
\noindent where we used the relations $RA = AR$ and $RB_l  = B_r$. It is clear from (\ref{eq9a}) and (\ref{eq10}) that the half-return maps $\mathbb{P}_r$ and $\mathbb{P}_l$ in (\ref{eq6}) and (\ref{eq7}) are identical in the reference frame of the right-handed juggler. This implies that the hybrid dynamics of the devil-stick between any two control actions can be described by a single return map if the change in reference frame of the right-handed juggler is incorporated in the dynamic model. This map, $\mathbb{P} : S \rightarrow S$, can be obtained by first rewriting the left-hand-sides of (\ref{eq6}), (\ref{eq9a}) and (\ref{eq10}) as follows:
\begin{subequations}\label{eq10-new}
\begin{alignat}{999}
RY_{\!Q}(t_{k+1}^{-}) &= A\, Y_{\!P}(t_{k}^{-}) + B_r,  &\quad k = 1, 3, 5, \cdots \cr
\Rightarrow \quad Y_{\!Q}(t_{k+1}^{-}) &= R\left[A\, Y_{\!P}(t_{k}^{-}) + B_r\right],  &\quad k = 1, 3, 5, \cdots \label{eq10-newa} \\
RY_{\!P}(t_{k+1}^{-}) &= A\, Y_{\!Q}(t_{k}^{-}) + B_r,  &\quad k = 2, 4, 6, \cdots \cr
\Rightarrow \quad Y_{\!P}(t_{k+1}^{-}) &= R\left[A\, Y_{\!Q}(t_{k}^{-}) + B_r\right],  &\quad k = 2, 4, 6, \cdots \label{eq10-newb}
\end{alignat}
\end{subequations}
Then, by accounting for the change in reference frame of the right-handed juggler after each control action, (\ref{eq10-newa}) and (\ref{eq10-newb}) can be combined into the following single equation which represents the return map $\mathbb{P}$:
\begin{equation*}
\bar{Y}(t_{k+1}^{-}) = R\left[A \bar{Y}(t_{k}^{-}) + B_r\right], \qquad k = 1, 2, 3, \cdots
\end{equation*}
\noindent where $\bar{Y}$ denotes the state vector $Y$ in the reference frame of the right-handed juggler. The above equation results in the following discrete-time equations:
\begin{subequations}\label{eq11}
\begin{align}
\omega(t_{k+1}^{-}) &= -\omega(t_{k}^{-}) - (I_k\, r_k/J) \label{eq11a} \\
h_x(t_{k+1}^{-}) &= - h_x(t_{k}^{-}) \!-\! \left[v_x(t_{k}^{-}) \!-\! (I_k/m)\sin\theta^* \right]\!\delta_k\!\label{eq11b} \\
v_x(t_{k+1}^{-}) &= -v_x(t_{k}^{-}) +  (I_k/m)\sin\theta^* \label{eq11c} \\
h_y(t_{k+1}^{-}) &=h_y(t_{k}^{-}) - (1/2) g\,\delta_k ^2 \cr
& \quad +\left[v_y(t_{k}^{-}) +  (I_k/m)\cos\theta^* \right]\delta_k \label{eq11d}\\
v_y(t_{k+1}^{-}) &= v_y(t_{k}^{-}) + (I_k/m)\cos\theta^*\!- \! g\, \delta_k  \label{eq11e}
\end{align}
\end{subequations}
\noindent where $\delta_k \triangleq (t_{k+1}^{-} - t_{k}^{-})$, $k = 1, 2, \cdots$, is the time of flight between two consecutive control actions. During this time duration, the devil-stick rotates by a net angle
$\pi-2\theta^*$. Since the angular velocity of the stick remains constant in the interval $[t_{k}^{+}, t_{k+1}^{-} ]$, $\delta_k$ is given as follows
\begin{align}\label{eq12}
\delta_k = \frac{\Delta\theta}{\omega(t_{k}^{-}) + (I_k\, r_k/J)}, \quad \Delta\theta \triangleq (\pi-2\theta^*)
\end{align}
\noindent The control design for juggling is presented next.\

\section{State Feedback Control Design}\label{sec4}
\subsection{Steady-State Dynamics}\label{sec4-1}
\noindent From the discussion in section \ref{sec3-5} it becomes clear that when the change of reference frame of the juggler is taken into account, the problem of juggling between the two distinct configurations $(\theta^*, h_x^*, h_y^*)$ and $(\pi - \theta^*, -h_x^*, h_y^*)$ is converted to the problem of juggling between identical configurations $(\theta^*, h_x^*, h_y^*)$ and $(\theta^*, h_x^*, h_y^*)$.  If the state variables at this configuration are denoted by
\begin{align}\label{eq13}
\bar{Y}^* \triangleq \left[\begin{matrix} \omega^* &h_x^* &v_x^* &h_y^* &v_y^* \end{matrix}\right]^T
\end{align}
\noindent then $\bar{Y}^* = \mathbb{P} (\bar{Y}^*)$ is a fixed point and (\ref{eq11}) and (\ref{eq12}) give
\begin{subequations}\label{eq14}
\begin{align}
\omega^* &= -\omega^* - (I_k^*\, r_k^*/J) \label{eq14a} \\
h_x^* &= - h_x^* \!-\! \left[v_x^* \!-\! (I_k^*/m)\sin\theta^* \right]\!\delta^*\!\label{eq14b} \\
v_x^* &= - v_x^* +  (I_k^*/m)\sin\theta^* \label{eq14c} \\
h_y^* &= h_y^* - (1/2) g\,{\delta^*}^2 +\left[v_y^* +  (I_k^*/m)\cos\theta^* \right]\delta^* \label{eq14d}\\
v_y^* &= v_y^* + (I_k^*/m)\cos\theta^*\!- \! g\, \delta^*  \label{eq14e}\\
\delta^* &= \frac{\Delta\theta}{\omega^* + (I_k^*\, r_k^*/J)} \label{14f}
\end{align}
\end{subequations}
\noindent where $I_k^*$, $r_k^*$ denote the steady-state values of the control inputs and $\delta^*$ denote the steady-state value of the time of flight. Since $h_y^*$ is eliminated from (\ref{eq14d}), (\ref{eq14}) represents six equations in seven unknowns, namely, $\omega^*$, $h_x^*$, $v_x^*$, $v_y^*$, $I^*$, $r^*$, and $\delta^*$. By choosing
$\delta^*$, the remaining six unknowns are obtained as follows:
\begin{equation}\label{eq15}
\begin{alignedat}{3}
\omega^* &= -\Delta\theta/\delta^*,\,\,\,\, &h_x^* &= g\,{\delta^*}^{2}\tan\theta^*/4\cr
v_x^* &= g\,\tan\theta^*\delta^*/2,&v_y^* &= -g\,\delta^*/2\cr
I^* &= mg\delta^*/\cos\theta^*, &r^* &= 2J\cos\theta^*\Delta\theta/(mg{\delta^*}^{2})
\end{alignedat}
\end{equation}
\noindent Since the point of application of the impulsive force must lie on the stick, $r^*$ in (\ref{eq15}) must satisfy $0 < r^* < \ell/2$. This imposes the following constraint of the value of
$\delta^*$:
\begin{align}\label{eq16}
\delta^*  > \bar\delta, \quad \bar\delta \triangleq 2\sqrt{\frac{J\cos\theta^*\Delta\theta}{mg\ell}}
\end{align}
\noindent It should be noted that for a given value of $\delta^*$, the value of $h_y^*$ is not unique.\

\subsection{Error Dynamics}
To converge the states to their desired values, we first define the discrete error variables:
\begin{equation}\label{eq17}
\begin{alignedat}{3}
\widetilde{\omega}(k) &\triangleq \omega(t_k^{-}) - \omega^* \cr
\widetilde{h}_x(k) &\triangleq h_x(t_k^{-}) - h_x^*,\,\, &\widetilde{v}_x(k) &\triangleq v_x(t_k^{-}) - v_x^* \cr
\widetilde{h}_y(k) &\triangleq h_y(t_k^{-}) - h_y^*,\,\, &\widetilde{v}_y(k) &\triangleq v_y(t_k^{-}) - v_y^*\cr
\widetilde{u}_1(k) &\triangleq (I_kr_k - I^*r^*)/J,\quad &\widetilde{u}_2(k) &\triangleq (I_k - I^*)/m
\end{alignedat}
\end{equation}
\noindent Using (\ref{eq11}) and (\ref{eq14a})-(\ref{eq14e}), the error dynamics can now be written as
\begin{subequations}\label{eq18}
\begin{align}
\widetilde{\omega}(k+1) = &-\widetilde{\omega}(k) -  \widetilde{u}_1(k) \label{eq18a} \\
\widetilde{h}_x(k+1) = &- \widetilde{h}_x(k) \!-\! \delta_k \widetilde{v}_x(k)  + \delta_k \sin\theta^* \,\widetilde{u}_2(k)\label{eq18b} \\
\widetilde{v}_x(k+1) = &-\widetilde{v}_x(k) +  \sin\theta^*\, \widetilde{u}_2(k) \label{eq18c} \\
\widetilde{h}_y(k+1) =  & \,\, \widetilde{h}_y(k) + \delta_k \widetilde{v}_y(k) + \delta_k \cos\theta^*\, \widetilde{u}_2(k) \cr
&\qquad\,\,\,\, + (g/2)\left[\delta_k \delta_k^* - \delta_k^2\right] \label{eq18d}\\
\widetilde{v}_y(k+1) = & \,\,\widetilde{v}_y(k) + \cos\theta^* \widetilde{u}_1(k) \!- \! g\, [\delta_k - \delta_k^*] \label{eq18e}
\end{align}
\end{subequations}
\noindent where $\delta_k$, defined in (\ref{eq12}), can be written in terms of the error variables as follows:
\begin{align}\label{eq19}
\delta_k = \frac{\Delta\theta\, \delta^*}{\left[\widetilde{\omega}(k) +\widetilde{u}_1(k)\right]\delta^* + \Delta\theta}
\end{align}
\noindent It is clear from (\ref{eq18}) and (\ref{eq19}) that the error dynamics is nonlinear. In the next section we present a partial control design that converts the nonlinear system into a linear system and simplifies the remaining control design.\

\subsection{Partial Control Design: Dead-Beat Control}
The error dynamics in (\ref{eq18}) involves two control inputs, namely, $\widetilde{u}_1(k)$ and $\widetilde{u}_2(k)$. The input $\widetilde{u}_1(k)$ appears only in (\ref{eq18a}). To this end, we first design $\widetilde{u}_1(k)$ as follows:
\begin{align}\label{eq20}
\widetilde{u}_1(k) &= -\widetilde{\omega}(k)
\end{align}
\noindent to guarantee dead-beat convergence of the error state $\widetilde{\omega}(k)$. Substitution of (\ref{eq20}) in (\ref{eq19}) yields $\delta_k = \delta^*$. Since, $\delta^*$ is user-defined and is a constant, the choice of control in (\ref{eq20}) is special as it transforms the remaining dynamics in (\ref{eq18b})-(\ref{eq18e}) into the linear system:
\begin{align}\label{eq21}
&z(k+1) = \mathcal{A}\,z(k) + \mathcal{B}\,\widetilde{u}_2(k)\cr
&z(k) \triangleq \left[\begin{matrix} \widetilde{h}_x(k) &\widetilde{v}_x(k) &\widetilde{h}_y(k) &\widetilde{v}_y(k) \end{matrix}\right]^T \cr
\mathcal{A} &\triangleq  \!\left[\begin{matrix} -1 &-\delta^* &0 &0 \cr 0 &-1 &0 &0 \cr 0 &0 &1 &\delta^* \cr 0 &0 &0 &1 \end{matrix}\right],\quad \mathcal{B} \triangleq  \!\left[\begin{matrix} \delta^*\sin\theta^* \cr \sin\theta^* \cr \delta^*\cos\theta^* \cr \cos\theta^* \end{matrix}\right]
\end{align}
\noindent It can be verified that the pair $(\mathcal{A}, \mathcal{B})$ is controllable since $\theta^* \in (0, \pi/2)$ and $\delta^* > 0$.\

\subsection{Residual Control Design}
The error dynamics in (\ref{eq21}) is linear and therefore the states can be converged to zero by simply designing a linear controller. However, it should be noted that the control input $\widetilde{u}_2(k)$ determines the value of $I_k$ which also appears in the dead-beat control design $\widetilde{u}_1(k)$ - see (\ref{eq17}). By using the values of $\widetilde{u}_2(k)$ from (\ref{eq17}) in (\ref{eq20}), we get:
\begin{align}\label{eq22}
r_k &= \left[I^*r^*-J \widetilde{\omega}(k)\right]/I_k
\end{align}
\noindent Since the point of application of impulsive force must lie of the stick, $r_k$ must satisfy $-\ell/2 < r_k < \ell/2$. By imposing this condition on the value of $r_k$ in (\ref{eq22}), we get the following constraints on the input $\widetilde{u}_2(k)$:
\begin{equation}
\begin{aligned}\label{eq23}
\widetilde{u}_2(k) &> \left[\,2I^*r^* - 2J\widetilde{\omega}(k) - I^*\ell\,\right]/(m\ell)\cr
\widetilde{u}_2(k) &> \left[\,-2I^*r^* + 2J\widetilde{\omega}(k) - I^*\ell\,\right]/(m\ell)
\end{aligned}
\end{equation}
\noindent Since $I^*$ and $r^*$ are both positive, as it can be seen from (\ref{eq15}), the inequalities in (\ref{eq23}) can be combined into the single inequality:
\begin{align}\label{eq24}
&\widetilde{u}_2(k) > \bar{a} + \bar{b}\mid\!\widetilde{\omega}(k)\!\mid \\
&\bar{a} \triangleq (2r^* - \ell)I^*/(m\ell), \quad \bar{b} \triangleq 2J/(m\ell) \notag
\end{align}
\noindent Since $\widetilde{u}_1(k)$ is dead-beat, $\widetilde{\omega}(k) = 0, k = 2, 3 \cdots$. Thus, (\ref{eq24}) can also be written as
\begin{equation}\label{eq25}
\begin{alignedat}{3}
\widetilde{u}_2(k) &> \bar{a} + \bar{b}\mid\!\widetilde{\omega}(k)\!\mid,\qquad  &&k = 1 \cr
\widetilde{u}_2(k) &> \bar{a},\quad &&k = 2, 3, \cdots
\end{alignedat}
\end{equation}

The input $\widetilde{u}_2(k)$ is designed using Linear Quadratic Regulator (LQR) and Model Predictive Control (MPC) methods. For an LQR design, the control minimizes the cost function
\begin{align}
J = \sum_{k=1}^\infty \left[z(k)^T Q\, z(k) + R\, \widetilde{u}_2^2(k)\right] \label{eq25-1}
\end{align}
where, $Q$ and $R$ are constant weighting matrices that can be chosen by trial and error to satisfy the constraints in (\ref{eq25}). The closed-form solution of the control input $\widetilde{u}_2(k)$ can be obtained by solving the Ricatti equation \cite{antsaklis}.\

For a receding horizon MPC design, the constraint in (\ref{eq25}) can be explicitly included in the optimization problem. In the MPC design\footnote{A detailed discussion of MPC design for discrete-time systems can be found in Chapters 1-3 in \cite{wang2009model}.}, it is necessary to calculate the predicted output with future control input as the adjusted variable. Since the current control input cannot affect the output at the same time for receding horizon control, the system dynamics must be represented in terms of the difference between the current and the predicted control input. To this end, we define the following variables based on the augmented state-space model\footnote{The augmented state-space model is controllable; this was verified using Theorem 1.2 in \cite{wang2009model}.} in \cite{wang2009model}:
\begin{equation}
\begin{aligned}
&\Delta u(k_i) \triangleq \widetilde{u}_2(k_i) - \widetilde{u}_2(k_i-1) \cr
&\Delta U_i \triangleq \left[\begin{matrix}\Delta u(k_i) \cr \Delta u(k_i + 1) \cr \vdots \cr \Delta u(k_i + N_c - 1)\end{matrix}\right],\,\,
Z_i \triangleq \left[\begin{matrix} z(k_i + 1 \mid k_i) \cr z(k_i + 2 \mid k_i) \cr \vdots \cr z(k_i + N_p \mid k_i)\end{matrix}\right]
\end{aligned}
\end{equation}
\noindent where $k_i$ is the current sampling instant, $z(k_i)$ is the state vector in (\ref{eq21}) measured at $k_i$, $N_c$ is the control horizon, $N_p$ is the prediction horizon, and $ z(k_i + m \mid k_i)$ is the predicted state variable at $k_i+m$ with state measurements $z(k_i)$.

We now construct the following $N$-step receding horizon optimal control problem:
\begin{align}
&\text{minimize}\,\, J = \sum_{i=1}^{N} \left[Z_i^T Z_i+ \Delta U_i^T \Delta U_i \right] \label{eq27} \\
&\text{subject to} \cr
&\begin{array}{lll} \label{eq28}
&z(k_i+1) = \mathcal{A}\,z(k_i) + \mathcal{B}\,\widetilde{u}_2(k_i) \\
&\widetilde{u}_2(k_i) > \bar{a} + \bar{b}\mid\!\widetilde{\omega}(1)\!\mid, \quad &i = 1 \\
&\widetilde{u}_2(k_i) > \bar{a}, \quad &i = 2, 3 \cdots, N
\end{array}
\end{align}
\noindent In every sampling period, the optimization problem determines the best control parameter $\Delta U_i$ that attempts to converge the sequence of states in $Z_i$ to zero. Although $\Delta U_i$ contains $N_c$ number of future control inputs, only the first entry is implemented as the actual control input. This optimization process is repeated using a more recent measurement of the states. It should be emphasized that the input constraint in (\ref{eq28}), namely, $\widetilde{u}_2(k_i) > \bar{a} + \bar{b}\mid\!\widetilde{\omega}(1)\!\mid$ is imposed only in the first optimization window. In subsequent optimization windows, the constraint is relaxed to $\widetilde{u}_2(k_i) > \bar{a}$. This is necessary for $\widetilde{u}_2$ to converge to zero since $\bar{a}$ is negative - see (\ref{eq24}), whereas $\bar{a} + \bar{b}\mid\!\widetilde{\omega}(1)\!\mid$ can assume positive values based on the initial value of $\widetilde\omega$.\
\begin{remark}
The control input $\widetilde{u}_2(k)$ is obtained as the numerical solution of the optimal control problem in (\ref{eq27}) and (\ref{eq28}). These inputs are applied at discrete time instants and the optimization solver is required to compute these inputs within the sampling time interval, which is equal to the time of flight $\delta^*$. Since $\delta^*$ is relatively large, there is sufficient time for the optimization solver to generate the solution. This, along with the fact that the input constraint can be explicitly considered in the problem formulation, makes MPC well-suited for this problem.
\end{remark}

\section{Simulation Results}\label{sec5}
\subsection{System Parameters and Initial Conditions}\label{sec5-1}
We present simulation results of both LQR- and MPC-based control designs. The physical parameters of the devil-stick are provided below in SI units:
\begin{align}\label{eq29}
m = 0.1,\quad  \ell = 0.5,  \quad J = 0.0021
\end{align}
Using these physical parameters and by choosing the values of $\theta^* = \pi/6$ rad and $\delta^* = 0.5$ sec, the steady-state values of state variables and control inputs are obtained from (\ref{eq15}) as
\begin{equation}\label{eq30}
\begin{alignedat}{4}
\omega^* &= -4.18\, \rm{rad/s} \quad &h_x^* &= 0.353\, \rm{m} \quad &v_x^* &= 1.414\, \rm{m/s}\cr
v_y^* &= -2.45\, \rm{m/s} \quad &I^* &= 0.565\, \rm{Ns} \quad &r^* &= 0.030\, \rm{m}
\end{alignedat}
\end{equation}
Since $h_y^*$ can be chosen arbitrarily, we chose
\begin{equation} \label{eq31}
h_y^* = 3.0\, \rm{m}
\end{equation}
At the initial time, we assume $\theta = \theta^* = \pi/6$ rad and the states variables (in SI units) are
\begin{equation}\label{eq32}
\begin{alignedat}{5}
\omega(0) &= 0, \quad &h_x(0) &= 0.53, \quad &v_x(0) &= 2.0 \cr
&&h_y(0) &= 1.0, &v_y(0) &= -2.0
\end{alignedat}
\end{equation}
\noindent For the physical parameters in (\ref{eq29}), steady-state values of the states in (\ref{eq30}) and (\ref{eq31}), and initial conditions in (\ref{eq32}), the control $\widetilde{u}_1(k)$ was chosen according to (\ref{eq20}). The control input $\widetilde{u}_2(k)$ was designed using LQR and MPC methods and simulation results are presented next.\

\subsection{Results for the LQR-based Design}
\begin{figure}[t!]
\centering
\psfrag{A}[][]{{\scriptsize{$k$}}}
\psfrag{B}[][]{{\scriptsize{$\omega(k)$ (rad/s)}}}
\psfrag{C}[][]{{\scriptsize{$h_x(k)$ (m)}}}
\psfrag{D}[][]{{\scriptsize{$h_y(k)$ (m)}}}
\psfrag{F}[][]{{\scriptsize{$v_x(k)$ (m/s)}}}
\psfrag{G}[][]{{\scriptsize{$v_y(k)$ (m/s)}}}
\psfrag{E}[][]{{\scriptsize{$E(k)$ (J)}}}
\includegraphics[width=0.90\hsize]{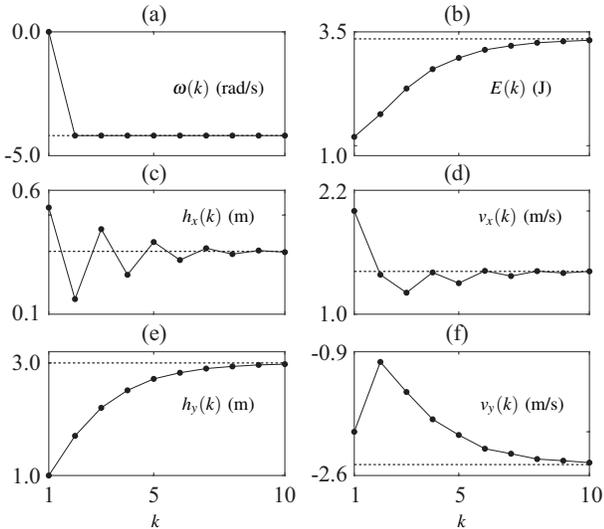}
\caption{State variables and total energy $E$ of the devil-stick at sampling instants $k$, $k = 1, 2, \cdots, 10$, for the LQR design.}
\label{Fig4}
\end{figure}
\begin{figure}[t!]
\centering
\psfrag{A}[][]{{\scriptsize{$k$}}}
\psfrag{B}[][]{{\scriptsize{$I_k$ (Ns)}}}
\psfrag{C}[][]{{\scriptsize{$r_k$ (m)}}}
\includegraphics[width=1.00\hsize]{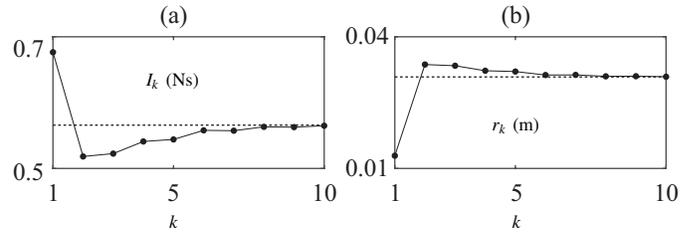}
\caption{Control inputs for the devil-stick at sampling instants $k$, $k = 1, 2, \cdots, 10$, for the LQR design.}
\label{Fig5}
\end{figure}
For the LQR design, the weight matrix $Q$ for the states was chosen to be the identity matrix and the control weight $R$ was chosen as $0.2$. The control was obtained as
\begin{align*}
\widetilde{u}_2(k) = F z(k), \quad F = \begin{bmatrix} -0.43 &-0.77 &0.43 &0.66\end{bmatrix}
\end{align*}
The simulation results are shown in Figs.\ref{Fig4}\footnote{It should be noted that the state variables are shown in the reference frame of the right-handed juggler.} and \ref{Fig5}. It can be seen from Fig.\ref{Fig4} (a) that the dead-beat control $\widetilde{u}_1(k)$ converges $\omega(k)$ to $\omega^*$ in one sampling interval. The control $\widetilde{u}_2(k)$ converges the remaining states to their steady-state values given in (\ref{eq30}) in approximately $k=10$ steps - see Figs.\ref{Fig4} (c)-(f). The control inputs $I_k$ and $r_k$ are shown in Figs.\ref{Fig5} (a) and (b). It can be seen that both control inputs converge to their steady-state values defined in (\ref{eq30}); also the control input $r_k$ remains well inside the constraint boundary $\mid\! r_k\! \mid < \ell/2$. The convergence of both the states and control inputs to their desired values imply that the devil-stick is juggled between two symmetric configurations. Since the magnitudes of $v_x$, $v_y$, $\omega_x$, and $h_y$ are the same in the two symmetric configurations, the total energy $E$ (kinetic plus potential) reaches a constant value at steady state - see Fig. \ref{Fig4}(b).\

\begin{remark}
The total energy of the devil-stick is the same at the symmetric configurations. Also, it is conserved during the flight phase. Therefore, in steady-state, the control inputs $I^*$ and $r^*$ do zero work on the system.
\end{remark}
\begin{figure}[t!]
\centering
\psfrag{A}[][]{{\scriptsize{$k$}}}
\psfrag{M}[][]{{\scriptsize{$I_k$ (Ns)}}}
\psfrag{N}[][]{{\scriptsize{$r_k $ (m)}}}
\psfrag{C}[][]{{\scriptsize{$h_x(k)$ (m)}}}
\psfrag{D}[][]{{\scriptsize{$h_y(k)$ (m)}}}
\psfrag{F}[][]{{\scriptsize{$v_x(k)$ (m/s)}}}
\psfrag{G}[][]{{\scriptsize{$v_y(k)$ (m/s)}}}
\includegraphics[width=0.90\hsize]{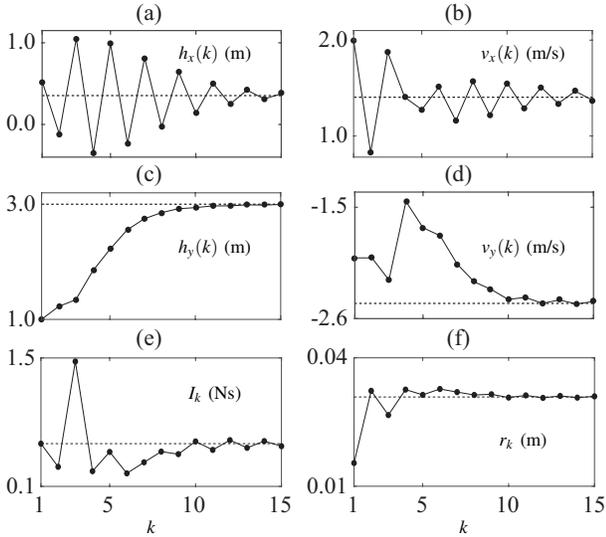}
\caption{State variables and control inputs of the devil-stick at sampling instants $k$, $k = 1, 2, \cdots, 15$, for the MPC design.}
\label{Fig6}
\end{figure}
\subsection{Results for the MPC-based Design}
\begin{figure}[t!]
\centering
\psfrag{A}[][]{{\scriptsize{$h_x$ (m)}}}
\psfrag{B}[][]{{\scriptsize{$h_y$ (m)}}}
\psfrag{C}[][]{{\scriptsize{$\theta^*$}}}
\includegraphics[width=1.00\hsize]{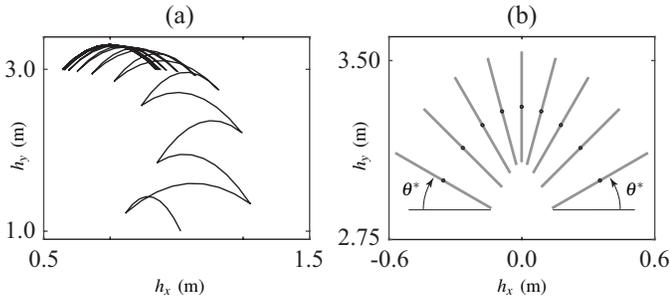}
\caption{(a) Trajectory of the center-of-mass from the initial configuration to steady-state and (b) symmetric configurations and seven intermediate configurations of the devil-stick in steady state for the MPC design.}
\label{Fig7}
\end{figure}

The control horizon, prediction horizon, and the number of steps were taken as
\begin{equation*}
N_c = 5, \quad N_p = 10, \quad N = 15
\end{equation*}
\noindent The MPC problem, defined by (\ref{eq27}) and (\ref{eq28}) were solved using quadratic programming in Matlab\footnote{The quadprog Matlab function was used.}. The state variables ${h}_x(k)$, ${h}_y(k)$,
${v}_x(k)$ and ${v}_y(k)$ and the control inputs $I_k$ and $r_k$ are shown in Fig.\ref{Fig6}. The state variable ${\omega}(k)$ is not shown as it converged to its desired value in one sampling interval by the dead-beat controller. Similar to the LQR design, the control input $r_k$ remains well inside the constraint boundary. The trajectory of the center-of-mass of the devil stick is shown in Fig.\ref{Fig7} (a); it starts from the initial configuration $(h_x, h_y) = (0.53, 1.00)$ and is eventually juggled between the symmetric coordinates $(h_x^*, h_y^*) = (0.353, 3.00)$ and $(-h_x^*, h_y^*) = (-0.353, 3.00)$ in steady state. Typically, $N$ is chosen to be large to guarantee convergence. For our system, the states rapidly converged to zero with $N=15$. In Fig.\ref{Fig7} (b), the devil-stick is shown at the two symmetric configurations where $\theta^* = \pi/6$ and several intermediate configurations that are equal time intervals apart.\

\begin{remark}
In both simulations, the stick rotates by an angle $(\pi - 2\theta^*)$ between two consecutive control inputs. This corresponds to ``top-only idle" juggling \cite{youtube1}. The controller is quite general and the stick can be controlled to rotate by $(q\pi - 2\theta^*)$, $q = 2, 3, \cdots$, by simply changing the definition of $\Delta\theta$ in (\ref{eq12}) from $\Delta\theta = (\pi - 2\theta^*)$ to $\Delta\theta = (q\pi - 2\theta^*)$. In other words, the stick can be made to flip multiple times in the flight phase, if desired. The ``flip-idle" in \cite{youtube1} corresponds to the case where $q = 2$.
\end{remark}

\section{Conclusion} \label{sec6}

Impulsive forces are applied intermittently for juggling a devil-stick between two symmetric configurations. A dynamic model of the devil-stick and a control design for the juggling task is presented here for the first time. The control inputs are the impulse of the impulsive force and its point of application on the stick. The control action is event-based and the inputs are applied only when the stick has the orientation of one of the two symmetric configurations. The dynamics of the devil-stick due to the control action and torque-free motion under gravity is described by two Poincar\'e sections; the symmetric configurations are fixed points of these sections. A coordinate transformation is used to exploit the symmetry and convert the problem into that of stabilization of a single fixed point. A dead-beat controller is designed to convert the nonlinear system into a controllable linear discrete-time system with input constraints. LQR and MPC methods are used to design the control inputs and achieve symmetric juggling. The LQR method has a closed-form solution and is easier to implement but requires trial and error to satisfy the input constraints. The MPC method has no closed-form solution as it is obtained by solving an optimization problem online. However, the optimization problem directly takes into account the input constraint. The computational cost of the MPC method, which can be a concern for many problems, is not a concern for the juggling problem since the time between consecutive control actions is relatively large.Simulation results validate both control designs and demonstrate non-prehensile manipulation solely using impulsive forces. Our future work will focus on robotic juggling; this includes design of experimental hardware, feedback compensation of energy losses due to inelastic collisions between the devil-stick and hand sticks, and motion planning and control of the robot end-effector for generating the impulsive forces designed by the control algorithms.\

\balance
\bibliographystyle{IEEEtran}
\bibliography{ref}

\end{document}